\documentclass[apjl]{emulateapj}

\shorttitle{The lowest gravity {\sc sd}B pulsator}
\shortauthors{\O stensen et al.}
\usepackage{graphicx}
\usepackage[dvips]{color}
\usepackage{txfonts}
\usepackage{natbib}
\usepackage{times}
\newcommand{\msol}{\ensuremath{\rm{M}_{\odot}}}

\newcommand{\mstar}{\ensuremath{\rm{M}_\star}}
\newcommand{\teff}{\ensuremath{T_{\rm{eff}}}}
\newcommand{\logg}{\ensuremath{\log g}}

\newcommand{\lheh}{\ensuremath{\log \left(N_{\mathrm{He}}/N_{\mathrm{H}}\right)}}
\newcommand{\pmra}{\ensuremath{\mu_{\alpha}}}
\newcommand{\pmdec}{\ensuremath{\mu_{\delta}}}
\newcommand{\twom}{{\sc 2mass}}
\newcommand{\galex}{{\sc galex}}
\newcommand{\uHz}{\ensuremath{\mu{\rm{Hz}}}}

\newcommand{\target}{J20136+0928}
\newcommand{\ltarget}{GALEX J201337.6+092801}

\begin{document}

\title{\ltarget: The lowest gravity subdwarf B pulsator}
\author{
   R.~H.~\O stensen\altaffilmark{1},
   P.~I.~P\'apics\altaffilmark{1},
   R.~Oreiro\altaffilmark{1,2},
   M.~D.~Reed\altaffilmark{3},
   A.~C.~Quint\altaffilmark{3},
   J.~T.~Gilker\altaffilmark{3},
   L.~L.~Hicks\altaffilmark{3}, \\
   A.~S.~Baran\altaffilmark{4,5},
   L.~Fox Machado\altaffilmark{6},
   T.~A.~Ottosen\altaffilmark{7,8} and
   J.~H.~Telting\altaffilmark{7}
}

\affil{$^1$ Instituut voor Sterenkunde, K.U.~Leuven, Celestijnenlaan 200D, 3001 Leuven, Belgium;
\textcolor{blue}{roy@ster.kuleuven.be}
}
\affil{$^2$ Instituto de Astrof\'isica de Andaluc\'ia, Glorieta de la Astronom\'ia s/n, 18008 Granada, Spain}
\affil{$^3$ Department of Physics, Astronomy, and Materials Science, Missouri State University, Springfield, MO 65804, USA}
\affil{$^4$ Mt.~Suhora Observatory, Cracow Pedagogical University, Podchorazych 2, 30-084 Krakow, Poland}
\affil{$^5$ Department of Physics and Astronomy, Iowa State University, Ames, IA 50011, USA}
\affil{$^6$ Observatorio Astron\'omico Nacional, 
   Universidad Nacional Aut\'onoma de M\'exico, 
   Ensenada, BC 22860, Mexico}
\affil{$^7$ Nordic Optical Telescope, 38700 Santa Cruz de La Palma, Spain}
\affil{$^8$ Department of Physics and Astronomy, Aarhus University, 8000 Aarhus C, Denmark}

\begin{abstract}
   We present the recent discovery of a new subdwarf B variable (sdBV),
   with an exceptionally low surface gravity. Our spectroscopy
   places it at \teff\,=\,32100$\pm$1000, \logg\,=\,5.15$\pm$0.20,
   and \lheh\,=\,--2.8$\pm$0.2. With a magnitude of $B$\,=\,12.0, it
   is the second brightest V361\,Hya star ever found.
   Photometry from three different
   observatories reveals a temporal spectrum with eleven clearly detected
   periods in the range 376 to 566\,s, and at least five more close to
   our detection limit. These periods are
   unusually long for the V361\,Hya class of short-period sdBV pulsators,
   but not unreasonable for $p$- and $g$-modes close to the radial fundamental,
   given its low surface gravity.
   Of the $\sim$50 short-period sdB pulsators known to date,
   only a single one has been found to have comparable spectroscopic
   parameters to \ltarget (\target, for short).
   This is the enigmatic high-amplitude pulsator V338\,Ser, and we
   conclude that \target\ is the second example of this rare subclass
   of sdB pulsators located well above the canonical extreme horizontal
   branch in the HR diagram.
\end{abstract}


\section{Introduction}

The subdwarf B (sdB) stars are generally known to be core-helium-burning
stars with inert hydrogen-dominated envelopes of very low mass.
This composition places them on the hot extension of the
horizontal branch,
the extreme horizontal branch \citep[EHB;][]{heber86,heber09}.
Their low-mass envelopes prevent them from reaching a second giant stage,
and instead, after their core helium is exhausted, they
expand only briefly, never exceeding more than about a third of a solar
radius, before contracting and passing on
towards higher effective temperatures and gravities.
Thus, an sdB star evolves into the hotter sdO population before reaching
degeneracy and the associated white dwarf (WD) cooling track \citep{dorman93}.

While the future evolution of sdB stars is unproblematic, exactly how
they lose their envelopes during their first red giant (RGB) stage is
more complicated.
If the star has a close low-mass companion, mass transfer on the tip of
the RGB will quickly pull it into the envelope.
As friction transfers angular momentum from the orbit to the envelope,
it will spin up and be ejected from the system.
If the core has become massive enough for helium to ignite ($\sim$0.47\,\msol),
the result will be a very close sdB+dM system. These binaries have spectacular
reflection effects, and a total of 14 such objects are described in the literature; 
13 are listed in \citet{for10} and the most recent one in \citet{twom1938}.
While such sdB+dM binaries make out only a tiny fraction of the total
sdB population, cases where the companion responsible for ejecting the
envelope is a low-mass WD, appear to be much more common.
\citet{maxted01} found that 21 of 36 stars in their radial velocity study are in
short period systems, most likely with WD companions.

If the companion is more massive than the subdwarf (at least at the
end of mass transfer), the orbit will have expanded to more than
a hundred days. Such orbits are hard to measure, but the companion
is easily detectable spectroscopically or from infrared excess.
\citet{napiwotzki04} found that more than a third of their sdB sample
show the spectroscopic signature of main sequence companions, while
\citet{reed04}, using \twom\ photometry, inferred that about half
of the sdBs in the field are likely to be of this type.
Thus, the vast majority of sdB stars in the field must be in binary systems. 
However, a few pulsating sdBs have been studied in such detail that any
companion close enough to have interacted during the RGB stage can be
ruled out. A possible explanation for such single sdB stars can be the merger
of two low-mass WDs.
These three scenarios are all investigated in detail in the binary population
study of \citet{han02,han03}. They find that while the first two scenarios
must have sdBs with masses close to the He-flash mass of 0.47\,\msol, the
merger scenario can produce sdBs with a broad distribution of masses
ranging from 0.4 to 0.7\,\msol.

\begin{figure*}
\includegraphics[width=\hsize]{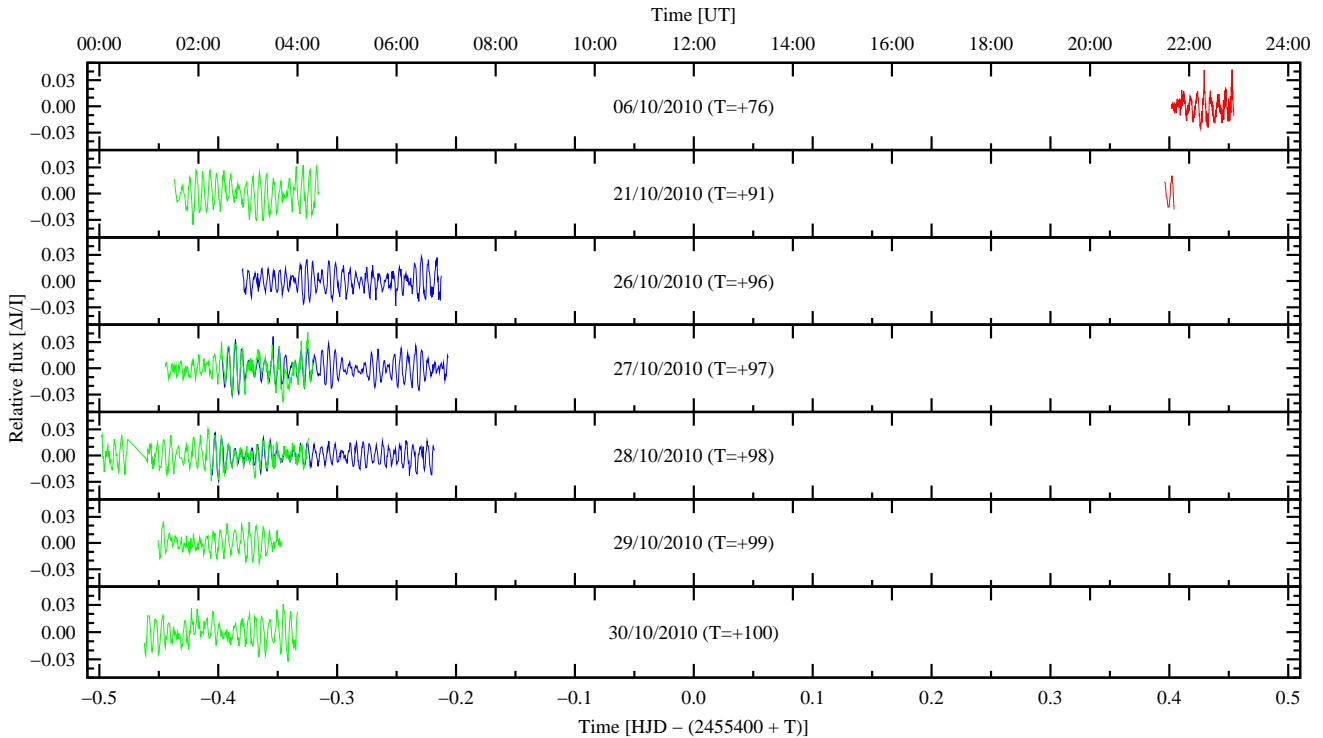}
\caption{
Light curves of \target. The top panel represents
the discovery light curve. 
In the online version the different data sets are shown in
different colors (Mercator; red, Baker; green, SPM; blue).
}
\label{fig:lc_all}
\end{figure*}

While most sdB stars form a band in the \teff--\logg\ diagram that is
consistent with a mass close to the canonical 0.47\,\msol, and various
envelope thicknesses up to the H-burning limit, a few exceptions are
found. One such is \object{V338\,Ser}, discovered to be a high
amplitude sdB pulsator by \citet{koen98_pg1605}.
A multi-site campaign devoted to this pulsator \citep{kilkenny99} found more
than fifty pulsation frequencies in the range 365--529\,s, which now defines the
long-period edge for the V361\,Hya group of short-period sdB variables (sdBVs).
Time-resolved spectroscopy of V338\,Ser was analysed by \citet{tillich07},
and most recently \citet{v338Ser} performed high-resolution time-resolved
spectroscopy.
Its unusually low gravity was discussed briefly by \citet{ostensen09},
proposing that it may be a core-He-burning EHB star with a mass larger
than the canonical, rather than a shell-He-burning post-EHB star
as suggested by \citet{koen98_pg1605}.
Asteroseismic solutions supporting this hypothesis were recently
presented by \citet{vanGrootel10_V338Ser}, implying a mass of
0.76\,\msol, and an envelope mass fraction of 0.2\%.

Since V338\,Ser remained a unique object in spite of a decade
of searches for V361\,Hya pulsators, we have recently been 
paying particular attention to candidates in this low-gravity region.
Here we present the discovery of pulsations in \target\
(Figure~\ref{fig:lc_all}), which has an
even lower surface gravity than V338\,Ser, and the lowest of any
sdB pulsator found to date.

\section{Target selection}

Since 2008 we have observed UV-excess targets selected based on
UV photometry from the {\em Galaxy Evolution Explorer} (\galex)
satellite \citep{GALEX}.
An estimate based on targets with NUV\,$<$\,14.0 and
FUV\,--\,B\,$<$\,+0.5 is that only about two-thirds
were already cataloged.
UV-excess objects in this cut include a host of interesting compact stars.
All types of pulsating WDs and hot subdwarf stars are
included, as are most types of cataclysmic variables
and planetary nebula nuclei. Our first cut based on \galex\ data
release 4 (GR4), contained 649 objects of which only about 400 had
even approximate classes in the literature. We started observing
targets from this list as a backup programme for various observing
runs, and have by now mostly completed the sample. Full details of
this survey will be given in a forthcoming paper. A similar survey
is described by \citet{vennes11}.
As we have accumulated spectroscopic
observations, new sdBV candidates have entered our observing
lists for photometric follow-up, and new discoveries are emerging.
The first result was the discovery of pulsations in \object{J08069+1527},
recently presented by \citet{baran11}, and establishing it as a being 
a high-amplitude member of the rare DW\,Lyn (hybrid) type of sdBVs.

The object presented here appears in \galex\ data release 6
(GR6) as \ltarget, with UV magnitudes
FUV\,=\,12.263(3), NUV\,=\,13.079(3).
Photographic magnitudes for this star from the {\sc nomad}\ database are
$B$\,=\,12.01, $V$\,=\,12.22, $R$\,=\,12.34, and from the \twom\
near-IR survey; 
$J$\,=\,13.00, $H$\,=\,13.11, $K$\,=\,13.23. The magnitudes
are all perfectly consistent with an sdB star that has no
main sequence companion.
\target\ also appears in the Tycho-2 catalog as TYC 1077-218-1,
where it has $B$\,=\,12.16, $V$\,=\,11.78.
The {\sc nomad}\ database also lists the target with a small but
significant proper motion (\pmra,\,\pmdec\,=\,12.1,\,2.8 mas/yr).

\begin{figure}
\centering
\includegraphics[width=6cm]{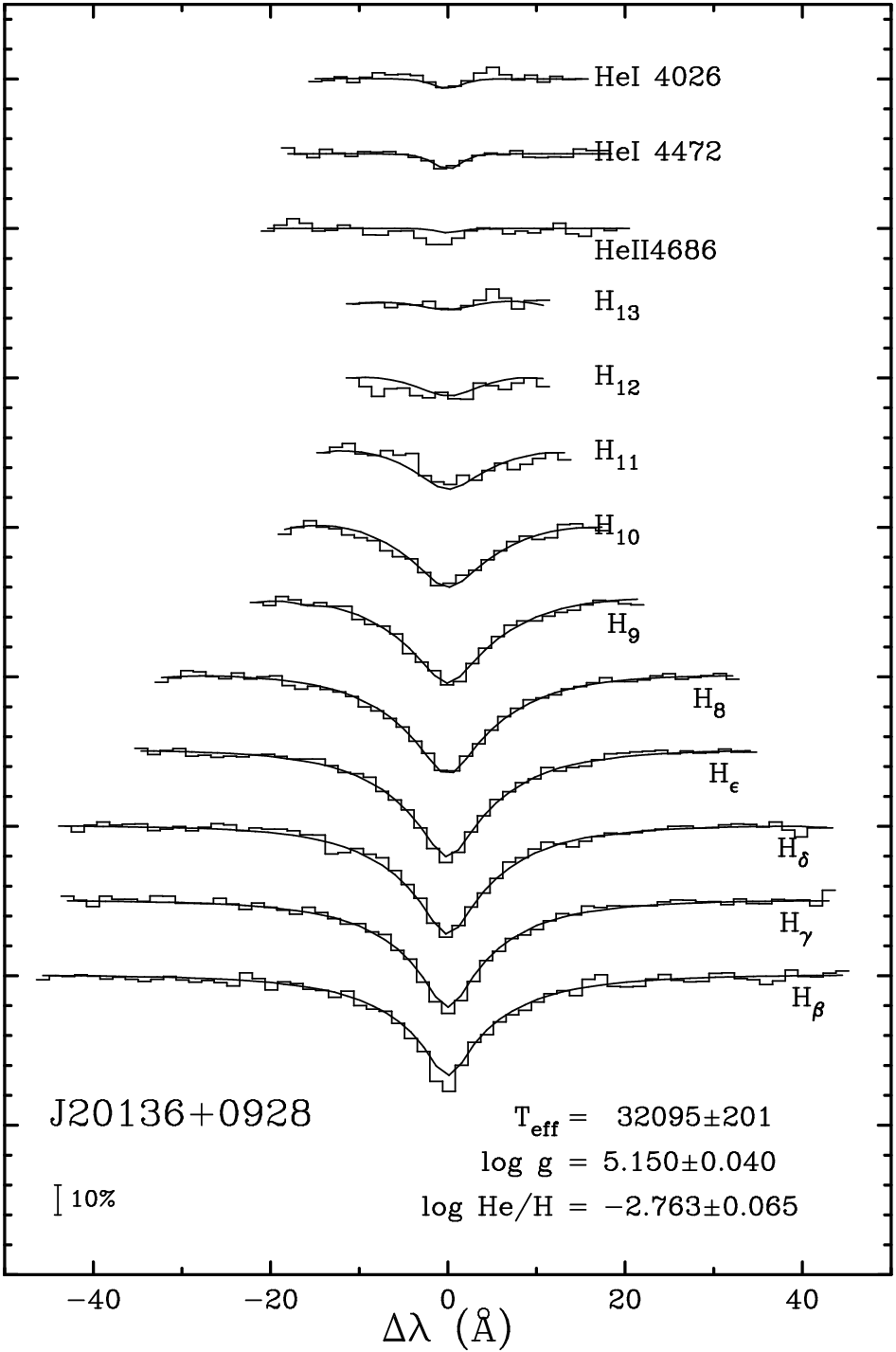}
\caption{LTE model fit of \target.}
\label{fig:tgfit}
\end{figure}


The target was observed spectroscopically during a run at the
Isaac Newton Telescope (INT) at 
Observatorio Roque de los Muchachos on the island of La Palma (Spain).
After the observing run we processed all the data and extracted the spectra
with {\sc iraf} in the regular way.  We fitted the hot subdwarfs using the
pure H+He LTE model grids of \citet{heber00}, and
we show the resulting fit for \target\ in Figure~\ref{fig:tgfit}.
The formal fitting errors (shown on the figure) do not account for
systematic effects inherent in the models, so we generously increase
the errors by a factor five when stating
\teff\,=\,32\,100$\pm$1000, \logg\,=\,5.15$\pm$0.20,
and \lheh\,=\,--2.8$\pm$0.2. Note that the typical shift of
when going from LTE to NLTE models is $\sim$+1000\,K, and
can become larger when metalicity effects are taken into account
\citep[see][and references therein]{heber09}.
However, the low surface gravity of this object is clear.


\begin{figure}
\centering
\includegraphics[width=8cm]{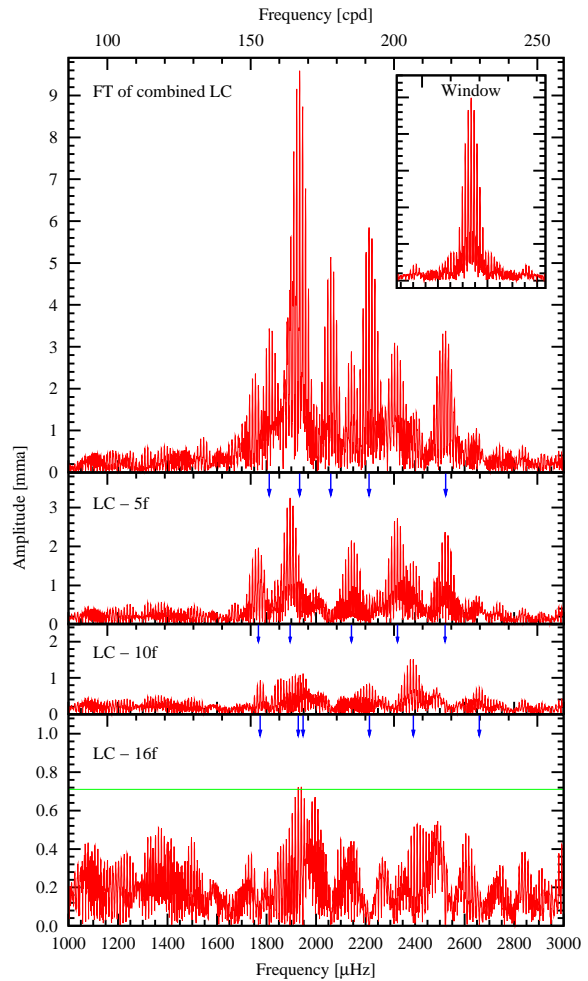}
\caption{FT of the five consecutive nights taken together
(top panel), and the prewhitening sequence after fitting
and subtracting five, ten and sixteen frequencies from
the original light curve. The window function is shown
as an inset to the top panel.
The arrows indicate the positions
of the frequencies identified and removed in each step.
The scale in the bottom panel has been expanded
to show the residual level, and the line at which we
stop is at 3.7 times the mean residual level.
}
\label{fig:ft_pw}
\end{figure}

\section{Photometric observations}

The discovery run was made on October 6, 2010 with the
1.2-m Mercator Telescope, sited just next to the INT.
Although the conditions were poor, the main period of $\sim$520\,s was
clearly seen, and recognised to be longer than normal for a V361\,Hya star.
Since it was already clear from the spectroscopy that the gravity of the
target was unusually low, the connection with V338\,Ser was immediately made,
and we urgently proceeded to confirm the discovery. 
As the target was already setting early in the
night, we pooled observing time at several telescopes in order to
collect enough photometry for a frequency analysis.

Photometric data were collected at three different sites within one
month of the discovery run. 
At the Mercator telescope we used the Merope CCD camera,
which was recently upgraded with a new E2V frame transfer CCD with
2048$\times$3074 illuminated pixels \citep{ostensen10}.
The observations were done with a Geneva-$B$ filter
($\lambda$\,=\,4201$\pm$286\,\AA), and a cadence period of $\sim$15 sec.
With the 0.4-m telescope at Baker Observatory (Missouri, USA) we used
a Photometrics RS-1340 CCD with a BG40 filter
($\lambda$\,=\,4750$\pm$1500), and a cadence of $\sim$30\,s.
We also obtained three consecutive nights with the 0.84-m telescope
at San Pedro M\'artir (SPM) observatory in Mexico.
The CCD camera is equipped with an E2V 4240 CCD, and we used
a Bessel-$B$ filter ($\lambda$\,=\,4350$\pm$980\,\AA),
and a cadence of 38\,s.

The light curves are all shown in Figure~\ref{fig:lc_all},
with the discovery run
on top.  Poor weather at Mercator left us unable to collect
a significant amount of data at our prime site.
Differential photometry was made with respect to the sum of three or four of the
brightest reference stars found within the field of view of the CCDs.
Most of this reference signal comes from TYC 1077-402-1, about 1.7' SW of
the target, which is comparable in brightness to \target.
The five bottom panels are consecutive nights, but the coverage is rather
poor due to the target only being visible for a few hours before setting.
On both October 27 and 28, there was a few hours of overlap between data
from Baker and SPM.
This allowed us to confirm that there is no significant difference in
amplitude between observations made with the two telescopes, in spite of
a significantly broader bandpass at Baker.
We only use the last five nights for the frequency analysis described
in the next section.

\newcommand{\tpm}{\,\ensuremath{\pm}\,}
\begin{table}
\caption[]{List of Frequencies}
\label{tbl:freqs}
\centering
\begin{tabular}{lcccr} \hline
   & Frequency & Period & Amplitude & Phase \\
ID & \multicolumn{1}{c}{(\uHz)}&  \multicolumn{1}{c}{(s)} &
     \multicolumn{1}{c}{(mma)} & \multicolumn{1}{c}{(s)} \\ \hline
$f_{9}$ & 1767.3\tpm0.3 & 565.84\tpm0.09 & 2.18\tpm0.38 &  194\tpm27 \\
$u_{15}$& 1774.5\tpm0.6 & 563.53\tpm0.19 & 1.28\tpm0.43 &   12\tpm56 \\
$f_{4}$ & 1810.7\tpm0.2 & 552.26\tpm0.06 & 3.91\tpm0.42 &  123\tpm18 \\
$f_{7}$ & 1894.8\tpm0.4 & 527.75\tpm0.11 & 3.07\tpm0.87 &  246\tpm40 \\
$u_{12}$& 1908.1\tpm0.7 & 524.08\tpm0.18 & 1.84\tpm0.71 &  344\tpm88 \\
$f_{1}$ & 1933.8\tpm0.1 & 517.12\tpm0.04 & 8.51\tpm0.82 &  139\tpm19 \\
$u_{11}$& 1946.9\tpm0.5 & 513.65\tpm0.14 & 2.11\tpm1.03 &  466\tpm54 \\
$f_{3}$ & 2059.3\tpm0.1 & 485.60\tpm0.03 & 4.59\tpm0.36 &   82\tpm11 \\
$f_{10}$& 2142.8\tpm0.3 & 466.67\tpm0.06 & 2.11\tpm0.36 &   97\tpm22 \\
$f_{2}$ & 2214.1\tpm0.3 & 451.66\tpm0.05 & 5.44\tpm1.15 &   70\tpm29 \\
$u_{14}$& 2215.2\tpm1.0 & 451.42\tpm0.19 & 1.45\tpm1.18 &   31\tpm107 \\
$f_{8}$ & 2328.8\tpm0.2 & 429.41\tpm0.04 & 2.88\tpm0.35 &  103\tpm15 \\
$f_{13}$& 2392.9\tpm0.4 & 417.91\tpm0.07 & 1.51\tpm0.35 &  405\tpm28 \\
$f_{6}$ & 2521.7\tpm0.4 & 396.56\tpm0.06 & 3.16\tpm0.99 &   31\tpm37 \\
$f_{5}$ & 2523.3\tpm0.3 & 396.30\tpm0.05 & 3.86\tpm0.99 &    0\tpm30 \\
$u_{16}$& 2659.5\tpm0.7 & 376.02\tpm0.10 & 0.75\tpm0.35 &   26\tpm50 \\
\hline
\end{tabular}\\
{\bf Notes.}---Phases are times of maxima since HJD 2455495.5.\\
Frequencies with large uncertainties are identified with $u$.
\end{table}

\begin{figure*}[t]
\centering
\includegraphics[width=12cm]{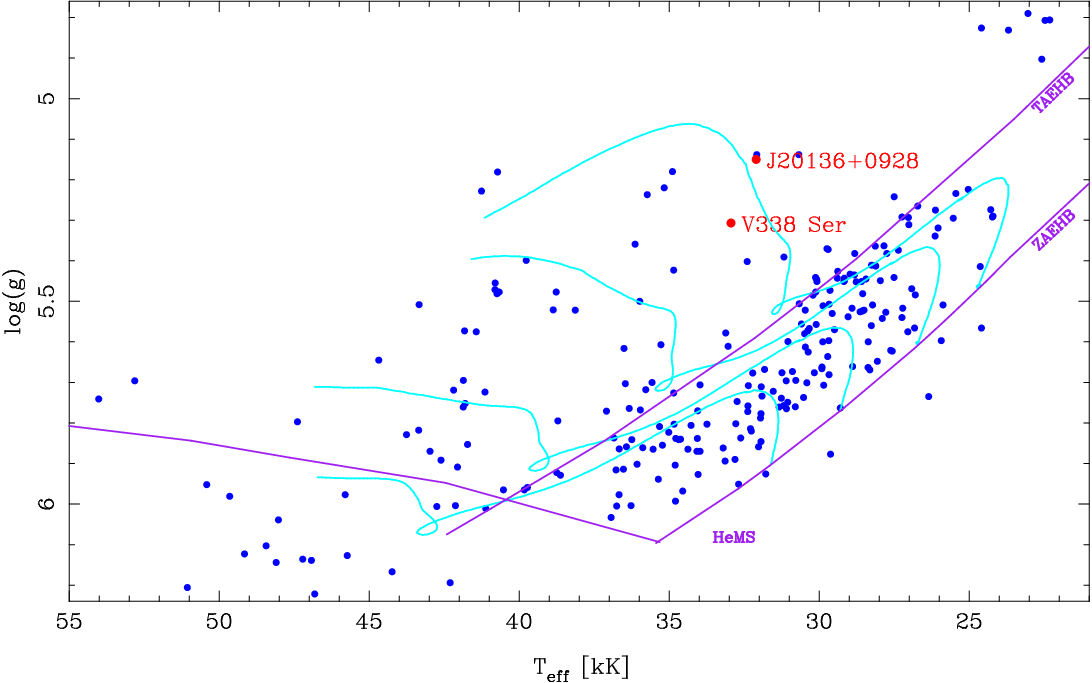}
\caption{The \teff--\logg\ diagram as observed by the Bok--Green survey
\citep{green08} with V338\,Ser highlighted, and \target\ added.
The evolutionary tracks are from \citet{kawaler05}.}
\label{fig:tgplot}
\end{figure*}

\section{Frequency analysis}

The Fourier transform (FT) of all runs longer than 1\,h
is far too short to resolve the complexity of the frequency
spectrum displayed by \target. Joining the five consecutive nights and
taking the FT of the combined dataset produces the complex frequency
spectrum with strong 1-day aliasing
shown in the top panel of Figure~\ref{fig:ft_pw}.
The tools used for the frequency determination were developed particularly for
multi-site data, and are described in \citet{vuckovic06}.
We proceed to prewhiten the most significant well-separated peaks, five or
six at a time, until we reach a level of 0.71\,mma (straight line in the
bottom panel of the same figure). This level is 4.2 times the noise level
when computed in the high-frequency region where no pulsational power
is seen (although we do detect the drive frequency of some of our 
telescopes), and 3.7 times the mean residual level of the 1000 to
3000\,\uHz\ region shown in the last panel.
In each step the most significant peaks are used as input for a non-linear
least-squares fitting procedure, and all 16 frequencies, amplitudes and phases
are left free in the final step.  Several frequencies close to the
most significant peaks do not converge to reliable values, and we consider
these as uncertain (IDs labeled $u$ in Table~\ref{tbl:freqs}).
It is not clear whether this is due to the poor coverage of our data, or caused
by real amplitude variability of the target. Concern can also be raised
about some frequencies that are barely resolved ({\em e.g.} $f_5$ and $f_6$).

\section{Discussion and Conclusions}

A new, bright sdBV with an exceptionally low gravity has been found in a sample
of UV-excess stars identified from \galex\ photometry.
In Figure~\ref{fig:tgplot} we show its
position with respect to V338\,Ser and the bulk of the sdB population, which
falls between the zero and terminal age EHB lines. \target\ is
located close to minimum \logg\ during the post-EHB stage of the uppermost
evolutionary track,
which corresponds to
a 0.470\,\msol~helium burning star with an H-envelope mass of 0.004\,\msol.
Evolutionary models of more massive EHB stars are not shown, but can
be found in \citet{han02}, from which one may infer that models with
a mass of $\sim$0.7\,\msol\ and an envelope of $\sim$0.01\,\mstar\ 
would approximate the position of \target\ reasonably well.
As mentioned in
the introduction, such an overmassive EHB star cannot form through the regular
binary mass-transfer scenarios that are responsible for producing the
bulk of the EHB stars, but can form through the merger of two helium
core WDs. Theoretical models for these scenarios,
combined with future observational efforts, may allow us to determine
the origins of both V338\,Ser and \target.

This discovery brings the total number of short period sdBV stars (both V361\,Hya
and hybrid DW\,Lyn stars) up to 52 (49 are summarised in \citealt{sdbnot}, one
was recently discovered in the field of the {\em Kepler} spacecraft by
\citealt{kawaler10a}, and the last one by \citealt{baran11}). 
The discovery of pulsations in a star with such a low gravity as \target\ is well
explained by the classical $\kappa$ mechanism for pulsations in sdBVs \citep{charpinet01}.
However, the exceptionally low gravity of \target\ is surprising in evolutionary
terms, and the presence of pulsations allows the mass and internal structure to
be explored.
The encouraging brightness of \target\ makes it an excellent target for
time-resolved spectroscopy studies, and arranging a multi-site campaign with small
telescopes when the target is visible throughout the night will be a priority for
the coming season.  By accurately determining the frequencies present in the star,
asteroseismology will be able to reliably determine its mass and internal
structure, and allow us to determine the most likely evolutionary
origins of this object.

\acknowledgements

The research leading to these results has received funding from the European
Research Council under the European Community's Seventh Framework Programme
(FP7/2007--2013)/ERC grant agreement N$^{\underline{\mathrm o}}$\,227224
({\sc prosperity}), as well as from the Research Council of K.U.Leuven grant
agreement GOA/2008/04.
RO acknowledges financial support from the Spanish grants AYA2009-08481-E,
AYA2010-14840 and AYA2009-14648-C02-02.
AB gratefully appreciates funding from Polish Ministry of Science and Higher
Education under project N$^{\underline{\mathrm o}}$\,554/MOB/2009/0.
ACQ, JTG and LLH were supported by the Missouri Space Grant, funded by NASA.
LFM acknowledges financial support from PAPIIT IN114309.

Based on observations made with the Mercator Telescope, operated on the
island of La Palma by the Flemish Community, at the Spanish Observatorio
del Roque de los Muchachos of the Instituto de Astrof\'{i}sica de Canarias.

\newcommand{\apjournal}[3]{#1, #2, #3}


\begin{thebibliography}{xx}

\bibitem[{{Baran} et~al.}(2011){Baran} et al.] {baran11}
{Baran}, A.~S., {Gilker}, J.~T., {Reed}, M.~D., {\O stensen}, R.~H.,
  {Telting}, J.~H., {Smolders} K., {Hicks} L.~\&\ {Oreiro} R. 
2011, \apjournal{\mnras}{in press}{doi:10.1111/j.1365-2966.2011.18356.x},
arXiv:1103.1600

\bibitem[{{Charpinet} et~al.}(2001)Charpinet et al.]{charpinet01}
{Charpinet}, C., {Fontaine}, G., \&\ {Brassard}, P.
2001, \apjournal{\pasp}{113}{775}

\bibitem[{{Dorman} {et~al.}(1993){Dorman}, {Rood}, \& {O'Connell}}]{dorman93}
{Dorman}, B., {Rood}, R.~T. \&\ {O'Connell}, R.~W. 
1993, \apjournal{\apj}{419}{596}

\bibitem[{{For} et~al.(2010)For et al.}]{for10}
{For}, B.~Q., et al.
2010, \apjournal{\apj}{708}{253},
arXiv:0911.2006

\bibitem[{{Green} et~al.(2008)Green et al.}]{green08}
{Green}, E.~M., {Fontaine}, G., {Hyde}, E.~A., {For}, B.~Q. \&\ {Chayer}, P. 
2008 in ASP Conf.~Ser.~392, Hot Subdwarf Stars and Related Objects,
ed.~U~{Heber}, C.~S {Jeffery} \&\ R~{Napiwotzki}
(San Francisco, CA: ASP), 75 

\bibitem[{{Han} et~al.(2003)Han et al.}]{han03}
{Han}, Z., {Podsiadlowski}, P., {Maxted}, P.~F.~L. \&\ {Marsh}, T.~R. 
2003, \apjournal{\mnras}{341}{669},
arXiv:astro-ph/0301380

\bibitem[{{Han} et~al.(2002)Han et al.}]{han02}
{Han}, Z., {Podsiadlowski}, P., {Maxted}., P.~F.~L., {Marsh}, T.~R. \&\ {Ivanova}, N. 
2002, \apjournal{\mnras} {336}{449},
arXiv:astro-ph/0206130

\bibitem[{{Heber}({1986})}]{heber86}
{Heber}, U.  1986, \apjournal{\aap} {155}{33}

\bibitem[{{Heber}({2009})}]{heber09}
{Heber}, U.  2009, \apjournal{\araa} {47}{211}

\bibitem[{{Heber} et~al.}(2000)Heber et al.]{heber00}
{Heber}, U., {Reid}, I.~N. \&\ {Werner}, K. 
2000, \apjournal{\aap}{363}{198}

\bibitem[{{Kawaler} \&\ {Hostler}}(2005)Kawaler, Hostler]{kawaler05}
{Kawaler}, S.~D. \&\ {Hostler}, S.~R. 
2005, \apjournal{\apj}{621}{432}

\bibitem[{{Kawaler} et~al.}(2010)Kawaler et al.]{kawaler10a}
{Kawaler}, S~D, et al.
2010, \apjournal{\mnras}{409}{1487},
arXiv:1008.2356

\bibitem[{{Kilkenny} et~al.}(1999)Kilkenny et al.]{kilkenny99}
{Kilkenny}, D, et al.
1999, \apjournal{\mnras} {303}{525}

\bibitem[{{Koen} et~al.}(1998)Koen et al.]{koen98_pg1605}
{Koen}, C., {O'Donoghue}, D., {Kilkenny}, D., {Lynas-Gray}, A.~E., {Marang}, F. \&\ {van Wyk} F. 
1998, \apjournal{\mnras} {296}{317}

\bibitem[{{Martin} et~al.}(2005)Martin et al.]{GALEX}
{Martin}, D.~C., et al.
2005 \apjournal{\apjl}{619}{L1}

\bibitem[{{Maxted} et~al.}(2001)Maxted et al.]{maxted01}
{Maxted}, P.~F.~L., {Heber}, U., {Marsh}, T.~R. \&\ {North}, R.~C. 
2001 \apjournal{\mnras}{326}{1391},
arXiv:astro-ph/0103342

\bibitem[{{Napiwotzki} et~al.}(2004)Napiwotzki et al.]{napiwotzki04}
{Napiwotzki}, R., et al.
2004 in ASP Conf.~Ser.~318, 
Spectroscopically and Spatially Resolving the Components of the Close Binary Stars
ed.~R.~W {Hilditch}, H~{Hensberge} \&\ K~{Pavlovski}
(San Francisco, CA: ASP), 402 

\bibitem[{{\O stensen}}(2009)Ostensen]{ostensen09}
{\O stensen}, R.~H.
2009, \apjournal{Communications in Asteroseismology}{159}{75},
arXiv:0901.1618

\bibitem[{{\O stensen}}(2010)Ostensen]{ostensen10}
{\O stensen}, R.~H.
2010, \apjournal{Astron.~Nachr.}{331}{1026},
arXiv:1010.3214

\bibitem[{{\O stensen} et al.(2010a)Ostensen et al.}]{twom1938}
{\O stensen}, R.~H., et al. 
2010a, \apjournal{\mnras}{408}{L51},
arXiv:1006.4267

\bibitem[{{\O stensen} et al.(2010b)Ostensen et al.}]{sdbnot}
{\O stensen}, R.~H., et al.
2010b, \apjournal{\aap}{513}{A6},
arXiv:1001.3657

\bibitem[{{\O stensen} et al.(2010c)Ostensen et al.}]{v338Ser}
{{\O}stensen}, R.~H., {Telting}, J.~H., {Oreiro}, R., {Heber}, U., {de Beck}, E. \&\ {Reed}, M. 
2010c, \apjournal{\apss}{329}{167}

\bibitem[{{Reed} \&\ {Stiening}(2004)Reed \&\ Stiening}]{reed04}
{Reed}, M.~D. \&\ {Stiening}, R.
2004, \apjournal{\pasp}{116}{506}

\bibitem[{{Tillich} et~al.(2007)Tillich et al.}]{tillich07}
{Tillich}, A., {Heber}, U., {O'Toole}, S.~J., {{\O}stensen}, R. \&\ {Schuh}, S.
2007, \apjournal{\aap}{473}{219},
arXiv:0707.0810

\bibitem[{{van Grootel} et~al.(2010)van Grootel et al.}]{vanGrootel10_V338Ser}
{van Grootel}, V., {Charpinet}, S., {Fontaine}, G. \&\ {Brassard}, P. 
2010, \apjournal{\apss}{329}{217}

\bibitem[{{Vennes} et~al.(2011)Vennes et al.}]{vennes11}
{Vennes}, S., {Kawka}, A. \&\ {N{\'e}meth}, P.
2011, \apjournal{\mnras}{410}{2095},
arXiv:1008.3823

\bibitem[{{Vu\v{c}kovi\'{c}} et~al.(2011)Vuckovic et al.}]{vuckovic06}
{Vu\v{c}kovi\'{c}}, M., et al.
2006, \apjournal{\apj}{646}{1230},
arXiv:astro-ph/0604330

\end{thebibliography}
\end{document}